\documentclass[singlespacing]{elsart}

\usepackage{graphicx}

\usepackage{amssymb}
\journal{Soft condensed matter}
\begin{document}

\begin{frontmatter}

\title{Discrete breathers in protein secondary structure}

\author{A.E. Sitnitsky},
\ead{sitnitsky@mail.knc.ru}

\address{Institute of Biochemistry and Biophysics, P.O.B. 30, Kazan
420111, Russia}

\begin{abstract}

The role of the rigidity of a peptide chain in its equilibrium
dynamics is investigated within a realistic model with stringent
microscopically derived coupling interaction potential and
effective on-site potential. The coupling interaction
characterizing the chain rigidity and the effective on-site
potentials are calculated for three main types of protein
secondary structure. The coupling interaction is found to be
surprisingly weak for all of them but different in character:
repulsive for $\alpha $-helix and anti-parallel $\beta $-sheet
structures and attractive for parallel $\beta $-sheet structure.
The effective on-site potential is found to be a hard one for
$\alpha $-helix and anti-parallel $\beta $-sheet and a soft one
for parallel $\beta $-sheet. In all three types of protein
secondary structures a stable zig-zag shape discrete breather (DB)
associated with the oscillations of torsional (dihedral) angles
can exist due to weakness of the coupling interaction. However,
since the absorption of far infrared radiation (IR) by proteins is
known to require the existence of rather long chains of hydrogen
bonds that takes place only in $\alpha $-helicies, then one can
conclude that the excitation of a DB in such a way is possible in
$\alpha $-helix and  seems to be hardly possible in $\beta
$-sheet structures. The interpretation of the recent experiments
of Xie et al. on far IR laser pulse spectroscopy of proteins is
suggested. The frequency of a DB in an the $\alpha $-helix is
obtained in the region of 115 $cm^{-1}$ in accordance with these
experiments.\\

\end{abstract}

\end{frontmatter}

\section{Introduction}
Nonlinearity opens new perspectives for solving the problem of
localization and storage of energy by biomolecules \cite{Dau92},
\cite{Pey95}, \cite{Pey00}. This problem is especially urgent
within the context of enzymatic catalysis. How does an enzyme
store and utilize the energy released at substrate binding? The
answer to this question is very poorly understood not only at
quantitative but even at qualitative level. By now it is clear
that dynamical contributions from linear modes into the mechanism
of enzyme action are negligible because of high friction in such
a condensed matter as protein interior leading to fast
dissipation of energy \cite{Fin02}. In contrast nonlinear systems
provide us with examples of surprising behavior: the higher the
friction is the slower the dissipation of energy proceeds
\cite{Gob02}! Though the generosity of this striking result is
under question one can hope that since proteins are very rich
structures for exhibiting nonlinear phenomena then some of them
may play a functional role in enzymatic reactions. Among
nonlinear phenomena in biomolecules the so called discrete
breather (DB) or else intrinsic localized mode was put in the
forefront during the last decade and even overshadowed such a
fashionable concept as soliton. However most results on DBs were
obtained within unrealistic models with toy potentials. That is
why the progress in application of this new paradigm to proteins
requires taking into account detailed information on protein
dynamics if one wish to pass from toy models to realistic ones.

Last years are marked by noticeable progress in revealing
subtleties of protein dynamics gained by infra-red (IR)
spectroscopy \cite{Wou02}, \cite{Wou01}, \cite{Ham98},
\cite{Ham00}, \cite{Xie00}, \cite{Xie01}, normal mode analysis
\cite{Yu03}, \cite{Lei02}, \cite{Lei01} and molecular-dynamics
simulations (see the recent reviews \cite{Kar02}, \cite{Kar00}
and refs. therein). On the one hand time-resolved 2D measurements
data suggest that in $\alpha $-helicies there are distributions
of the torsional (dihedral) angles ($\varphi$ and $\psi$) around
their average values with a width of 20 degrees and that the
conformation fluctuates on a time scale of picoseconds
\cite{Wou01} confirming previous molecular-dynamics studies. On
the other hand the experiments of Xie et al. \cite{Xie00},
\cite{Xie01} demonstrate that very long-lived oscillations can
exist in the $\alpha $-helix proteins (myoglobin,
bacteriorhodopsin) but not in the $\beta $-sheet one (photoactive
yellow protein) when excited by a far IR laser pulse. The
oscillations are observed in the Amide-I band ($\omega  \sim
1500$ cm$^{-1})$ \cite{Xie00} and more vividly in the
low-frequency band $\omega \sim100$ cm$^{-1}$ \cite{Xie01}. The
absorption of radiation by proteins in the latter case is commonly
attributed to the excitation of longitudinal and transverse modes
of one-dimensional hydrogen bonded chains in $\alpha $-helices
\cite{Col95}. The normal mode analysis of myoglobin \cite{Yu03},
\cite{Lei02}, \cite{Lei01} testifies that the anharmonic decay
rate of higher frequency localized normal modes, calculated by
perturbation theory, is typically nearly independent of
temperature, consistent with results of pump-probe studies on
myoglobin. In fact, the long-lived vibrations in the Amide-I
region of myoglobin \cite{Xie00} are not all that long-lived,
about 15 ps, so that this is really a predictable result. Indeed,
Leitner \cite{Lei01} and Yu and Leitner \cite{Yu03} have
presented calculations of anharmonic decay rates in the Amide-I
region of myoglobin, revealing that vibrational lifetimes of
order 1-10 ps are quite likely and unsurprising. On the other
hand, the results of \cite{Xie01} remain a puzzle. In this study,
long-lived excitations of over 500 ps are found near 115
$cm^{-1}$. Though there are no direct comparisons for
bacteriorhodopsin, the anharmonic decay rates in myoglobin near
100 $cm^{-1}$ computed by Yu and Leitner are two orders of
magnitude faster than this.

According to the conjecture of the authors of \cite{Xie00},
\cite{Xie01} their result suggests that a breather can be created
in an $\alpha $-helix. Another interpretation of these experiment
within the framework of Davydov's soliton concept is also
suggested \cite{D02}. The latter utilizes the possibility of
longitudinal mode excitation at the absorption of far IR
radiation by proteins. However this interpretation encounters with
difficulties at the point emphasized in \cite{Xie00} that $\alpha
$-helices in myoglobin or bacteriorhodopsin are too short for
such soliton could efficiently propagate in them. In this chapter
the hypothesis is put forward  that the long-lived oscillations in
$\alpha $-helicies are the fluctuations of the torsional angles
of peptide groups around their equilibrium values. The aim of this
chapter is to derive the possibility of the existence of a DB in
polypeptide chain from its microscopic model and to interpret the
phenomena observed the papers \cite{Xie00}, \cite{Xie01}. Our
approach emphasizes another possibility as opposed to the paper
\cite{D02}: the excitation of transverse mode of one-dimensional
hydrogen bonded chains in $\alpha $-helices at the absorption of
far IR radiation by proteins is actually that of oscillations of
the peptide groups around their equilibrium positions in peptide
chain (the oxygen and nitrogen atoms oscillate transverse to the
chain of peptide groups). This type of motion is shown to enable
the existence of DBs in $\alpha $-helices. Thus the DB in the
peptide chain of $\alpha $-helices can provide sustaining the
long-lived oscillations excited by the absorption in the
low-frequency band $\omega \sim$ 100 $cm^{-1}$.

DBs or else intrinsic localized modes (time periodic spatially
localized oscillations with significant amplitudes of several
units in a chain of weakly coupled non-linear oscillators while
others are at rest or oscillate with negligible amplitudes)
discovered by Sievers and Takeno in 1988 \cite{Siev88} and proved
by MacKay and Aubry \cite{Mac94} to be structurally stable are
well understood and commonly appreciated as generic phenomenon in
nature at present (see \cite{Fla98} and refs. therein). The key
ingredient for the existence of a DB is the requirement of weak
coupling interaction between adjacent nonlinear oscillators
\cite{Mac94}, \cite{Fla98}. DBs have become a new and very
fruitful paradigm in nonlinear physics. The concept of DB has
been applying to DNA dynamics for a long time \cite{Pey89},
\cite{Dau92}, \cite{Pey95}, \cite{Pey00}, \cite{Cue02},
\cite{Qas04}. Its application to protein dynamics is damped by a
point of view (see e.g. \cite{Pey00}) that proteins are less
capable than DNA to exhibiting DBs because they are much less
regular structures. However there are fragments of secondary
structures in proteins that are highly regular. In \cite{Arch02}
a model for $\alpha $-helix protein is considered that exhibits
DB existence in the chains of peptide groups connected by
hydrogen bonds in the spirit of Davydov's model for a soliton. In
this chapter a model is constructed that pursues the same goal
but for the peptide groups connected by C - C bonds in secondary
structures of the backbone. Another distinction from
\cite{Arch02} and from all other paper on DBs in biomolecules is
that we deal with realistic effective on-site potentials and
coupling interaction potentials rather than with model toy ones.
These potentials are derived from a stringent microscopic model
of the polypeptide chain and thus the present work seems to be a
step to ab initio calculation of DBs in biomolecules. There are
few analytical models treating non-linear mechanics of peptide
chain backbone \cite{Pey95}, \cite{Zor97}, \cite{Zor99},
\cite{Mer96}. The reason for transparent lack of activity on
analytical modeling of peptide chain backbone mechanics seems to
stem from meager direct experimental evidence for nonlinear
excitations in it. Nevertheless the experiment \cite{Xie01} can
be considered as that and motivates the present attempt to
construct a simple and tractable but at the same time
microscopically stringent model of polypeptide backbone dynamics.

\section{The Hamiltonian of the model}

A schematic picture of a polypeptide chain with all designations
used further is presented in Fig.1. The mutual orientation of two
adjacent peptide groups is characterized by the torsional
(dihedral) angles $\varphi _{i}$ and $\psi _{i}$. The angle
$\varphi _{i}$ characterizes the rotation round the bond
$C_{i}^{\alpha }- N_{i}$ and that $\psi _{i}$ characterizes the
rotation round the bond $C_{i}^{\alpha }- C^{'}_{i}$. Further we
consider the equilibrium dynamics in protein secondary structures
in which the equilibrium values of the torsional angles are the
same for all peptide groups (they are listed at the end of the
next Sec.). We consider small deviations $\varphi _{i}(t)$ and
$\psi _{i}(t)$ of the torsional angles from their equilibrium
values ($\varphi _{i}= \varphi _{i}^{0}+\varphi _{i}(t)$ and $\psi
_{i}=\psi _{i}^{0}+\psi _{i}(t)$) with $\left|{\varphi _{i}
(t)}\right| \le 20^{\circ}$ and $\left|{\psi _{i}(t)}\right| \le
20^{\circ}$. At such amplitudes of the deviations the hydrogen
bonds confining the peptide group in the secondary structure are
not broken \cite{Fin02}. Finally we consider a peculiar type of
motion $\psi _{i-1}(t)=\varphi _{i}(t)$. The latter means that
the peptide group rotates as a whole respective some effective
axis $\sigma$ passing through the center of the C$-$N bond
parallel to the bonds $C_{i}^{\alpha }- C^{'}_{i}$ and
$N_{i+1}-C^{\alpha} _{i+1}$ that are assumed to be approximately
parallel ($\angle C_{i}^{\alpha }-C^{'}_{i}- N_{i+1} =113^\circ$
and $\angle C^{'}_{i}- N_{i+1}- C^{\alpha} _{i+1} =123^\circ$
\cite{Can80}). This type of motion is stipulated by the fact that
the peptide group is a planar rigid structure \cite{Fin02},
\cite{Can80}. For the sake of uniformity of designations we
further denote $x _{i}=2\varphi _{i}(t)=2\psi _{i-1}(t)$ (see
Fig.2) and thus
\begin{equation}
\label{eq1} \varphi _{i} = \varphi _{i}^{0} + x _{i}/2 ;\quad
\quad \quad \psi _{i-1} = \psi _{i-1}^{0} + x _{i}/2
\end{equation}
The moment of inertia of the peptide group relative to the axis
$\sigma$ can be easily calculated and is $I \approx 7.34*10^{-39}
g*cm^{2}$. The Hamiltonian of the polypeptide chain in our model
with nearest neighbor interactions is
\begin{equation}
\label{eq2} H = \sum\limits_{i} {\left\{ \frac{I}{2}\left(
\frac{dx _{i} }{dt} \right)^{2} +U_{loc}(x_{i})+U(x_{i}; x_{i+1})
\right\}}
\end{equation}
Here $U_{loc}(x_{i})$ includes interactions defining the local
potential of the peptide group (namely hydrogen bonds, the
so-called torsional potentials and the van der Waals interaction
of covalently non-bonded atoms of the peptide group with the atoms
of adjacent side chains) defining its separate motion while
$U(x_{i}; x_{i+1})$ includes the coupling interactions (namely the
van der Waals interaction of covalently non-bonded atoms of
adjacent peptide groups i and i+1 and their electrostatic
interaction) intermixing the motions of these groups and leading
to the rigidity of the peptide chain. It should be stressed that
the latter potential also contributes into the separate motion of
the peptide groups. We can define the effective on-site potential
for such motion as
\begin{equation}
\label{eq3} V_{eff}(x_{i})=U_{loc}(x_{i})+U(x_{i}; 0)+U(0;x_{i})
\end{equation}
and the coupling interaction potential as
\begin{equation}
\label{eq4} U(x_{i}; x_{i+1})=U^{vdw}(x_{i};
x_{i+1})+U^{el}(x_{i}; x_{i+1})
\end{equation}
Thus the equation of motion is
\begin{equation}
\label{eq5}
I\frac{d^2x_i}{dt^2}=-\frac{dU_{loc}(x_i)}{dx_i}-\frac{dU(x_{i-1};
x_{i})}{dx_i}-\frac{dU(x_{i}; x_{i+1})}{dx_i}
\end{equation}
In the following two sections we consider in details the functions
$U_{loc}(x_i)$ and $U(x_{i}; x_{i+1})$. At doing it we will
freely pass back and forth between the variables according to the
rule (\ref{eq1}) which can be rewritten as
\begin{equation}
\label{eq6} \bigtriangleup\varphi _{i} =  x _{i}/2 ;\quad \quad
\quad  \bigtriangleup\psi _{i} =  x _{i+1}/2
\end{equation}

\section{Coupling interaction defining the rigidity of a polypeptide chain}

Both interactions in (\ref{eq4}) intermixing the motions of the
adjacent peptide groups and contributing to the coupling
interaction are described by the central potentials
$W^{vdw}_{mn}(R_{mn}(\varphi _{i};\psi _{i}))$ and
$W^{el}_{mn}(R_{mn}(\varphi _{i};\psi _{i}))$ between the atom
$A_{n}$ of the i-th peptide group and the atom $A_{m}$ of the
i+1-th one with $R_{mn}$ being the distance between the atoms
$A_{n}$ and $A_{m}$. The electrostatic potential is
\begin{equation}
\label{eq7} W_{mn}^{el} \left( {R_{mn}} \right) = \frac{{q_{m}
q_{n}} }{{\varepsilon R_{mn}}}
\end{equation}
\noindent where $q_{m}$ and $q_{n}$ are partial charges on the
atoms $A_{m}$ and $A_{n}$ respectively ($q(N)=-0.28e; q(H)=0.28e;
q(O)=-0.39e; q(C)=0.39e$ where $e=4.8*10^{-10}$ CGS \cite{Can80})
and $\varepsilon $ is the dielectric constant which for protein
interior should be better conceived as some adjustable parameter
($ \approx 2 \div 10$ with 3.5 being commonly accepted value
close to high frequency permeability of peptides) \cite{Can80},
\cite{Fin02}. In some studies $\varepsilon $ is supposed to be
solvent dependent and chosen, e.g., 4 for $CCl_4$, 6$\div$7 for
$CHCl_3$ and 10 for $H_2O$ \cite{Das87}, \cite{Pop97}. For the
van der Waals potential one can choose any of the numerous forms
suggested in the literature, e.g., the Lennard-Jones one (6-12) or
the Buchinghem one (6-exp). In this chapter we use for numerical
estimates the former one
\begin{equation}
\label{eq8} W^{vdw}\left( {R_{mn}}  \right) = - \frac{{A_{mn}}
}{{R_{mn}^{6}} } + \frac{{B_{mn}} }{{R_{mn}^{12}} }
\end{equation}
\noindent with the set of the well known parameters of Scott and
Scheraga (other sets were also verified and found to give similar
results).\\

The distance $R_{mn}$ as a function of the angles $\varphi _{i}$
and $\psi _{i}$ is
\[
 R_{mn} (\varphi _i ,\psi _i ) = \{ p_m^2  + r_n^2  +
 \]
 \[
 2p_m r_n [\sin \theta (\cos \gamma _m \sin \alpha _n \cos \varphi _i -
 \sin \gamma _m \cos \alpha _n \cos \psi _i  ) -
\]
 \begin{equation}
\label{eq9}
 \sin \gamma _m \sin \alpha _n (\cos \psi _i \cos \varphi _i \cos \theta  -
 \sin \psi _i \sin \varphi _i ) - \cos \gamma _m \cos \alpha _n \cos \theta ]\} ^{1/2}
\end{equation}

The interaction potential for both van der Waals and
electrostatic interactions has the form
\begin{equation}
\label{eq10}
 U^{\left\{ {\scriptstyle vdw \hfill \atop
  \scriptstyle el \hfill} \right\}} (x _i ;x _{i + 1} ) =
  \sum\limits_{mn} {W^{\left\{ {\scriptstyle vdw \hfill \atop
  \scriptstyle el \hfill} \right\}} } (R_{mn} (x _i ;x _{i + 1} ))
\end{equation}
\noindent where the summation in n is over atoms in the i-th
peptide group and that in m is over atoms in the i+1-th one.

In what follows we work with full realistic coupling interaction
potential described above. However for making use of suggestive
analogies with the known literature results obtained on toy
models we calculate the so called coupling constant which is a
key characteristic of the truncated coupling interaction
potential. Expanding the potential we obtain that to the leading
order in the terms $x _{i}\ll 1$ and $x _{i+1}\ll 1$ the rigidity
of the peptide chain is determined by the term of the potential
which intermixes the motion of the adjacent peptide groups
\begin{equation}
\label{eq11} U^{mix}\left( {x _{i} ;x _{i + 1}}  \right) \approx
(-K)x _{i}x _{i + 1}
\end{equation}
where the coupling constant is
\[
-K =-(K^{vdw} + K^{el}) =
\]
\begin{equation}
\label{eq12}\frac{{\partial ^{2}U^{vdw}\left( {x _{i} ;x _{i +
1}}  \right)}}{{\partial x _{i} \partial x _{i + 1}} }_{\left| {x
_{i} = 0;x _{i + 1} = 0} \right.} + \frac{{\partial
^{2}U^{el}\left( {x _{i} ;x _{i + 1}} \right)}}{{\partial x _{i}
\partial x _{i + 1}} }_{\left| {x _{i} = 0;x _{i + 1} = 0} \right.}
\end{equation}

The results of calculations of the contributions into the
coupling constant for different types of protein secondary
structures are as follows:

a). $\alpha $-helix (right) $\varphi _{i}^{0}=-57^{\circ}$; $\psi
_{i}^{0}=-47^{\circ}$
\cite{Can80}:$-K^{vdw}\approx4.24\cdot10^{-15}$ erg;
$-K^{el}\approx2.4\cdot10^{-15}$ erg at $\varepsilon =3.5$;
$-K\approx6.24\cdot10^{-15}$ erg at $\varepsilon =3.5$.

b). anti-parallel $\beta $-sheet $\varphi _{i}^{0}=-139^{\circ}$;
$\psi _{i}^{0} =135^{\circ}$
\cite{Can80}:$-K^{vdw}\approx-5.0\cdot10^{-16}$ erg;
$-K^{el}\approx2.64\cdot10^{-15}$ erg at $\varepsilon =3.5$;
$-K\approx2.14\cdot10^{-15}$ erg at $\varepsilon =3.5$.

c). parallel $\beta $-sheet $\varphi _{i}^{0}=-119^{\circ}$; $\psi
_{i}^{0}=113^{\circ}$\cite{Can80}:
$-K^{vdw}\approx-8.3\cdot10^{-15}$ erg;
$-K^{el}\approx2.96\cdot10^{-15}$ erg at $\varepsilon =3.5$;
$-K\approx-5.34\cdot10^{-15}$ erg at $\varepsilon =3.5$.

\section{Local potential of the peptide group}
The local potential $U_{loc}(x_i)$ is composed as follows
\begin{equation}
\label{eq13}
U_{loc}(x_i)={U}_{sc}(x_i)+{U}_{hb}(x_i)+{U}_{tors}(x_i)
\end{equation}
Here ${U}_{sc}(x_i)$ is the energy of van der Waals interactions
of the atoms of the i-th peptide group with those of the side
chains $R_i$ and $R_{i-1}$ and also with the atoms $H^i_{\alpha}$
and $H^{i-1}_{\alpha}$,
${U}_{hb}(x_i)={U}^{(1)}_{hb}(x_i)+{U}^{(2)}_{hb}(x_i)$ is the
energy of two hydrogen bonds of the i-th peptide group and
${U}_{tors}(x_i)={U}_{tors}^{\varphi}(\varphi
_{i})+{U}_{tors}^{\psi}(\psi _{i-1})$ is the energy of the
torsional potentials for the rotation of the angles $\varphi
_{i}$ and $\psi _{i-1}$. For the latter we take the usual form
\begin{equation}
\label{eq14} {U}_{tors}^{\varphi}(\varphi
_{i})=E_{\varphi}(1+\cos3\varphi _{i})
\end{equation}
\begin{equation}
\label{eq15}{U}_{tors}^{\psi}(\psi _{i})=E_{\psi}(1+\cos3\psi
_{i})
\end{equation}
with $E_{\varphi}\approx$1 kkal/mol and $E_{\psi}\approx$1
kkal/mol. In all further simulations we choose the side chain to
be Alanine ($R_i$ is $C^i_{\beta}(H)_3$) and initially find the
value of the angle $\chi_1$ (see Fig.1) to minimize the energy.
For the ${U}^{(j)}_{hb}(x_i)$ (j=1,2) we take the potential
\cite{Das87}, \cite{Pop97}
\begin{equation}
\label{eq16}
U^{(j)}_{hb}(x_i)=D{\left(1-exp[-n(r(x_i)-r_0)]\right)}^2-D
\end{equation}
where n=3 \AA$^{-1}$, $r_0$=1.8 \AA, D$\approx 4 \div 6$ kkal/mol
at $\varepsilon =3.5$ is an adjustable parameter of the hydrogen
bond energy with D=5 kkal/mol being a conventional value
\cite{Fin02} (in \cite{Pop97} D is assumed to be a decreasing
value at increasing $\varepsilon$ with D=0.5 kkal/mol at
$\varepsilon =10$) and $r(x_i)$ is the current length of the
hydrogen bond. The dependence of the latter on $x_i$ requires
some simple and straightforward but very cumbersome trigonometry
which we omit here to save room. It should be only mentioned that
this dependence can be expressed via the sets of the adjustable
parameters {P, L, $\eta$} and {R, L, $\kappa$} (see Fig.1). Here
L is the length of the hydrogen bond at equilibrium ($x_i$=0) (in
the literature the values from 1.8 \AA till 2.5 \AA are figured),
$\eta$ and $\kappa$ are the angles characterizing the extent of
linearity of the hydrogen bonds and finally R and P are the
distances from crossing of the perpendiculars of the $O_{i-1}$
and $H_i$ atoms on the axis $\sigma$ to their hydrogen bond
partners respectively at equilibrium ($x_i$=0). In the present
model we assume the partners to have fixed positions. Since in
the $\alpha $-helix the partners come from third peptide groups
along the chain from a given one this assumption imposes a
restriction on the size of a DB which can be considered, namely
the assumption can be justified for DBs comprising no more than 5
sites (peptide groups). The elimination of the restriction for
the $\alpha $-helix would actually mean taking into account the
long range interaction $U_{hb}(x_i;x_{i\pm3})$ along the peptide
chain. Such complication is left for future work and in this
chapter we exemplify the existence of a DB in the $\alpha $-helix
by a 3 - site one. For $\beta $-sheets the restriction can be
discarded because the partners usually belong to very distant
peptide groups from a given one.

The adjustable parameters are chosen to satisfy the following set
of requirements: 1. For all three types of protein secondary
structure the effective on-site potential $V_{eff}(x_{i})$ must
have a minimum at equilibrium ($x_i=0$) because the structures are
steady stable ones. 2. For the anti-parallel $\beta $-sheet the
angles $\eta$ and $\kappa$ are known to be very close to
$180^{\circ}$ \cite{Can80}. 3. For the  $\alpha $-helix the
spectroscopic frequency
\begin{equation}
\label{eq17}
(1/\lambda)_{sp}=\frac{\sqrt{V''_{eff}(x_i=0)/I}}{2\pi c}
\end{equation}
must be 115 $cm^{-1}$ for the model to give interpretation of the
experiment of \cite{Xie01}. Here $\lambda$ is a wavelength, c is
the light speed and the dash denotes a derivative in $x_i$. These
requirements are satisfied,  e.g.,  at the following values of the
common adjustable parameters: $\varepsilon =3.5$, D=5.7 kkal/mol,
$E_{\varphi}=1$ kkal/mol and $E_{\psi}=1$ kkal/mol and the
following values of the particular adjustable parameters:

a). $\alpha $-helix (right) $\varphi _{i}^{0}=-57^{\circ}$; $\psi
_{i}^{0}=-47^{\circ}$ \cite{Can80}: L=2.026 \AA, R=3.68 \AA,
P=3.34 \AA, $\eta=170^{\circ}$, $\kappa=170^{\circ}$.

b). anti-parallel $\beta $-sheet $\varphi _{i}^{0}=-139^{\circ}$;
$\psi _{i}^{0} =135^{\circ}$ \cite{Can80}: L=2 \AA,  R=3.61 \AA,
P=2.82 \AA, $\eta=180^{\circ}$, $\kappa=180^{\circ}$.

c). parallel $\beta $-sheet $\varphi _{i}^{0}=-119^{\circ}$; $\psi
_{i}^{0} =113^{\circ}$ \cite{Can80}: L=1.887 \AA,  R=3.55 \AA,
P=3.03 \AA, $\eta=170^{\circ}$, $\kappa=170^{\circ}$.

It seems plausible that the found values of the parameters are not
unique and some variation of them compatible with the
requirements mentioned above is possible but we could find no
reasonable ways to it. However full exploration of the parameter
space of the present model is a daunting task and we can not
exclude the possibility that the found values of the parameters
differ from reality. Nevertheless considering them as most
reasonable and coinciding with the conventional figures we
further restrict ourselves by dealing with them only.

\section{Results}
In what follows we use dimensionless variables. We define the
frequency
\begin{equation}
\label{eq18} \omega=\sqrt{\frac{V''_{eff}(x_i=0)}{I}}
\end{equation}
and measure time in the units of $\omega^{-1}$ so that the
dimensionless time is
\begin{equation}
\label{eq19} \tau=t\omega
\end{equation}
Also we measure energy in the units of $I\omega^2$ and denote the
dimensionless coupling constant $\rho $
\begin{equation}
\label{eq20} \rho = \frac{K}{I\omega ^2}
\end{equation}
The results of the investigation of the effective on-site
potential and the coupling interaction potential can be summarized
as follows:

a). $\alpha $-helix $((1/\lambda)_{sp}$=114.75 $cm^{-1}):
\rho=-0.00193; V'''_{eff}(x_i=0)>0$ (for $x\leq0.3$ the expansion
is $\frac{V_{eff}(x)}{I\omega ^2}\approx0.5 x^2 + 0.017 x^3 -
0.235 x^4 + 0.6 x^5 - 1.936 x^6 - 3.8 x^7 + 8.437 x^8 + 14.455
x^9 + 0.094 x^{10} - 42.98 x^{11} + 47.6 x^{12}+...)$. The
effective on-site potential is presented in Fig.3.

b). anti-parallel $\beta $-sheet $((1/\lambda)_{sp}$=84.34
$cm^{-1}):\rho=-0.00115; V'''_{eff}(x_i=0)>0$ (for $x\leq0.3$ the
expansion is $\frac{V_{eff}(x)}{I\omega ^2}\approx0.5 x^2 + 0.004
x^3 - 0.035x^4 - 4.22 x^6 - 0.0001 x^7 + 19.08 x^8 + 0.21 x^{10}
+ 148.844 x^{12}+...)$.

c). parallel $\beta $-sheet $((1/\lambda)_{sp}$=85.49
$cm^{-1}):\rho=0.0028; V'''_{eff}(x_i=0)<0$ (for $x\leq0.3$ the
expansion is $\frac{V_{eff}(x)}{I\omega ^2}\approx0.5 x^2 - 0.3
x^3 + 2.3x^4 + 3. x^5 - 15.9 x^6 - 14.6 x^7 + 57.65 x^8 + 52.4
x^9 + 0.64 x^{10} - 158.2 x^{11} + 403.3 x^{12}+...)$.

These results mean the following. For all three structures the
coupling interaction is surprisingly weak providing good
conditions for the existence of a DB. For the $\alpha $-helix and
the anti-parallel $\beta $-sheet the effective on-site potential
is hard (goes more steep than a harmonic one) and the coupling
interaction is repulsive (a non-zero value of a peptide group
displacement tends to increase the values of the neighboring
peptide groups displacements with the opposite sign). In
accordance with the theorems of the paper \cite{Arch02} a stable
DB must be of zig-zag shape (peptide groups undergo out of phase
oscillations). For the parallel $\beta $-sheet the effective
on-site potential is soft (goes less steep than a harmonic one)
and the coupling interaction is attractive (a non-zero value of a
peptide group displacement tends to increase the values of the
neighboring peptide groups displacements with the same sign). In
accordance with the theorems of the paper \cite{Arch02} a stable
DB must also be of zig-zag shape. Thus for no type of protein
secondary structure a bell shape DB (peptide groups undergo in
phase oscillations) can be stable.

In what follows we restrict ourselves by considering the case of a
DB in the $\alpha $-helix because only for it the experimental
data suggest such a phenomenon \cite{Xie01} (see the beginning of
the Discussion). The set of difference equations (\ref{eq5}) can
not be solved analytically and is analyzed by direct numerical
integration for 25 coupled equations (motivated by the fact that
25 is the average number of peptide groups in $\alpha $-helicies
of bacteriorhodopsin used in the experiment \cite{Xie01}). We use
fixed-end boundary conditions $x_1(t)=x_{25}(t)=0$ that seem to be
more appropriate for the situation in proteins than commonly used
periodic ones. We find as an example a 3-site DB. To prove that a
localized and stable object can be obtained in the $\alpha
$-helix we resort to the usual procedure \cite{Fla98}. We define
the discrete per site energy in 3 lattice sites around the 13-th
central one
\begin{equation}
\label{eq21} E(3)=\frac{1}{3}\sum_{l=12}^{14}e_l
\end{equation}
where $e_l$ is the individual symmetrized local site energy
density
\begin{equation}
\label{eq22}
e_l=\frac{1}{2}(\frac{dx_l}{d\tau})^2+\frac{U_{loc}(x_{l})}{I\omega
^2}+\frac{1}{2I\omega ^2}(U(x_{l-1}; x_{l})+U(x_{l}; x_{l+1}))
\end{equation}
The existence of a localized and stable zig-zag shape DB is
exemplified in Fig.4 by the fact that the energy E(3) remains
essentially constant with time and in Fig.5 by an explicit
picture. It should be stressed that for our realistic potentials
it is rather difficult to switch out the coupling interaction
without modifying the effective on-site potential. Thus the
commonly accepted way to construct a DB starting from an
uncoupled (or else anticontinuous) limit is inapplicable in our
case. That is why, instead, we find the initial configuration to
fall into a DB from the very beginning at a given coupling
interaction for the $\alpha $-helix calculated in Sec.3. In Fig.4
and Fig.5 this configuration is chosen to be $x _{13}(0)=0.3$;
$x_{12}(0)=x_{14}(0)=-0.025$ while $x_{i} (0)=0$ for all other i
and zero velocity at each site. The chosen amplitude $x_{13}=0.3$
corresponds to the deviations of the torsional angles
$\bigtriangleup\varphi _{13} = \bigtriangleup\psi _{12} = x
_{13}/2\approx10^{\circ}$ from their equilibrium values.

\section{Discussion}
The results obtained testify that a localized and stable zig-zag
shape DB can exist in all three types of protein secondary
structure. However experimental data which can be interpreted
with the help of the DB concept are available only for $\alpha
$-helix proteins but not $\beta $-sheet ones \cite{Xie00},
\cite{Xie01}. In our opinion the reason for this is in the way of
excitation of long-lived oscillations in this experiments rather
than in the propensities of different types of protein secondary
structure to sustaining DBs. The absorption of far IR laser pulse
used in \cite{Xie01} by proteins requires long one-dimensional
hydrogen bonded chains \cite{Col95} that take place in $\alpha
$-helicies but are absent in $\beta $-sheets.

The revealed hardness of the effective on-site potential for the
$\alpha $-helix and the anti-parallel $\beta $-sheet is in
contrast with a widely spread point of view that such potential in
biomolecules is mainly determined by hydrogen bonds and that is
why should be a soft-type one \cite{Pey00}, \cite{Arch02}. In our
case detailed account of all interactions leads to a contrary
result.

The energy of the 3-site DB obtained for the $\alpha $-helix is
$3E(3)I\omega ^2\approx2\cdot10^{-12}erg\approx40 k_BT$ at T=300
$K^{\circ}$. This value means that at room temperatures large
amplitude DBs
($\bigtriangleup\varphi=\bigtriangleup\psi\approx10^{\circ}$) can
hardly be spontaneously excited by thermal fluctuations. However
the latter may not be true for DBs with smaller amplitudes and
here it seems timely to touch upon the question of influence of
temperature on the existence of DBs in protein secondary
structures. DBs are known to remain essentially stationary and
localized and to be very long-lived at zero temperature. At
non-zero temperatures thermal fluctuations tend to lead to DB
motion and more rapid decay \cite{Pey95}, \cite{Pey00},
\cite{Rei02}. Within the framework of our interpretation of the
the experiment of \cite{Xie01} with the help of a DB concept it
should be recalled that the observed there long-lived
oscillations in $\alpha $-helicies of bacteriorhodopsin excited by
far IR laser pulse have a life-time of order of 500 ps
accommodating approximately 1500 vibrations. The decay of a DB
can be formally modeled by introducing a dissipative term
$-\gamma\frac{dx_i(t)}{dt}$ in the left hand side of the equation
of motion (\ref{eq5}) and corresponding white noise $\zeta_i(t)$
in the right hand one. Here the friction coefficient $\gamma$
must be related with the white noise correlation function via the
fluctuation-dissipation theorem
$<\zeta_i(t)\zeta_i(0)>=2k_BT\gamma\delta(t)\delta_{ij}$ where
$k_B$ is the Boltzman constant, T is the temperature, $\delta(t)$
is Dirac $\delta$-function and $\delta_{ij}$ is the Kronecker
symbol. However such formal way of introducing dissipation does
not reveal the microscopic origin of the friction coefficient
$\gamma$. To do it for the case of peptide chain dynamics is a
difficult task which is out of the scope of this chapter. Here we
can afford ourselves only some preliminary speculations on this
subject. There are different contributions into the mechanism of
friction that can be conceivable for the peptide chain rotational
dynamics in protein interior. The most obvious hydrodynamic
damping by interaction with solvent seems to be of minor
importance because at considered deviations of the torsional
angles $\leq 10^{\circ}$ the linear displacements are much less
than the diameter of a water molecule (see also \cite{Xie01} and
refs. therein). In our opinion the dominant contribution into
dissipation originates from the fact that the rotation of the
peptide group at angle x round the axis $\sigma$ requires some
translational displacement of axis points where the atoms $N_i$
and $C^{'}_{i-1}$ are initially situated (see Fig.2) that leads to
transvers distortions of the backbone (actually this is the
excitation of transvers linear phonon modes). Such process
consumes energy leading to an acoustic damping mechanism. At a
macroscopic level this dissipation mechanism can be described by
the so called "Landau damping" $\gamma_{ac}\propto \omega^2$
\cite{Lan01}. However such phenomenological description requires
support from microscopic consideration that is beyond our
knowledge at present. Moreover the situation with dissipation is
not so unambiguous as one can conceive from above. There are
testimonies that in some cases of nonlinear lattices the energy
relaxation from a DB first accelerates with the increase of the
friction coefficient from zero value but then becomes slower and
slower \cite{Gob02}! It is not clear at present whether this
surprising result is generic or it is model dependent (e.g.,
dependent on the type of the thermostat used). In any case it
opens interesting perspectives for application of the DB concept
to the problem of localized vibrational energy storage in
enzymes. In this connection the coincidence of the obtained value
for the DB energy ($\approx40 k_BT$ at room temperature) with
usual activation energies of enzymatic reactions is rather
intriguing. One can imagine oneself that in secondary structures
of enzymes there DBs can be excited at substrate binding and be
long-lived enough to play some functional role as the sources of
energy localization and storage utilizing in such a way the
binding energy. For instance a DB may serve as the so called "rate
promoting vibration" (RPV) for enzymatic reactions involving
enviromentally assisted hydrogen tunneling. The RPV acquires more
and more experimental testimonies \cite{Scru99}, \cite{Sut00},
\cite{Sut02}, \cite{Ant02}, \cite{Kna02} but its physical origin
still remains mysterious. However the estimated dominant peaks in
the spectral densities of the RPV indicate motions on the 150
$cm^{-1}$ frequency scale \cite{Ant02} that is rather close to
the obtained frequency 115 $cm^{-1}$ of the DB in the
$\alpha$-helix. Here we refrain from further discussing this
speculation.

\section{Conclusion}
The conclusions of this chapter are summarized as follows. The
dynamics of a peptide chain is considered within a realistic model
with stringent microscopically derived coupling interaction
potential and effective on-site potential. The coupling
interaction is found to be surprisingly weak for all three main
types of protein secondary structure but different in character:
repulsive for $\alpha $-helix and anti-parallel $\beta $-sheet
structures and attractive for parallel $\beta $-sheet structure.
The effective on-site potential is found to be a hard one for
$\alpha $-helix and anti-parallel $\beta $-sheet and a soft one
for parallel $\beta $-sheet. In all three types of protein
secondary structure a stable zig-zag shape discrete breather
associated with the oscillations of torsional (dihedral) angles
can exist due to the weakness of the coupling interaction.\\

Acknowledgements. The author is grateful to Prof. V.D. Fedotov,
R.Kh. Kurbanov and B.Z. Idiyatullin for helpful discussions.

\newpage
\begin{figure}
\begin{center}
\includegraphics* [width=\textwidth]{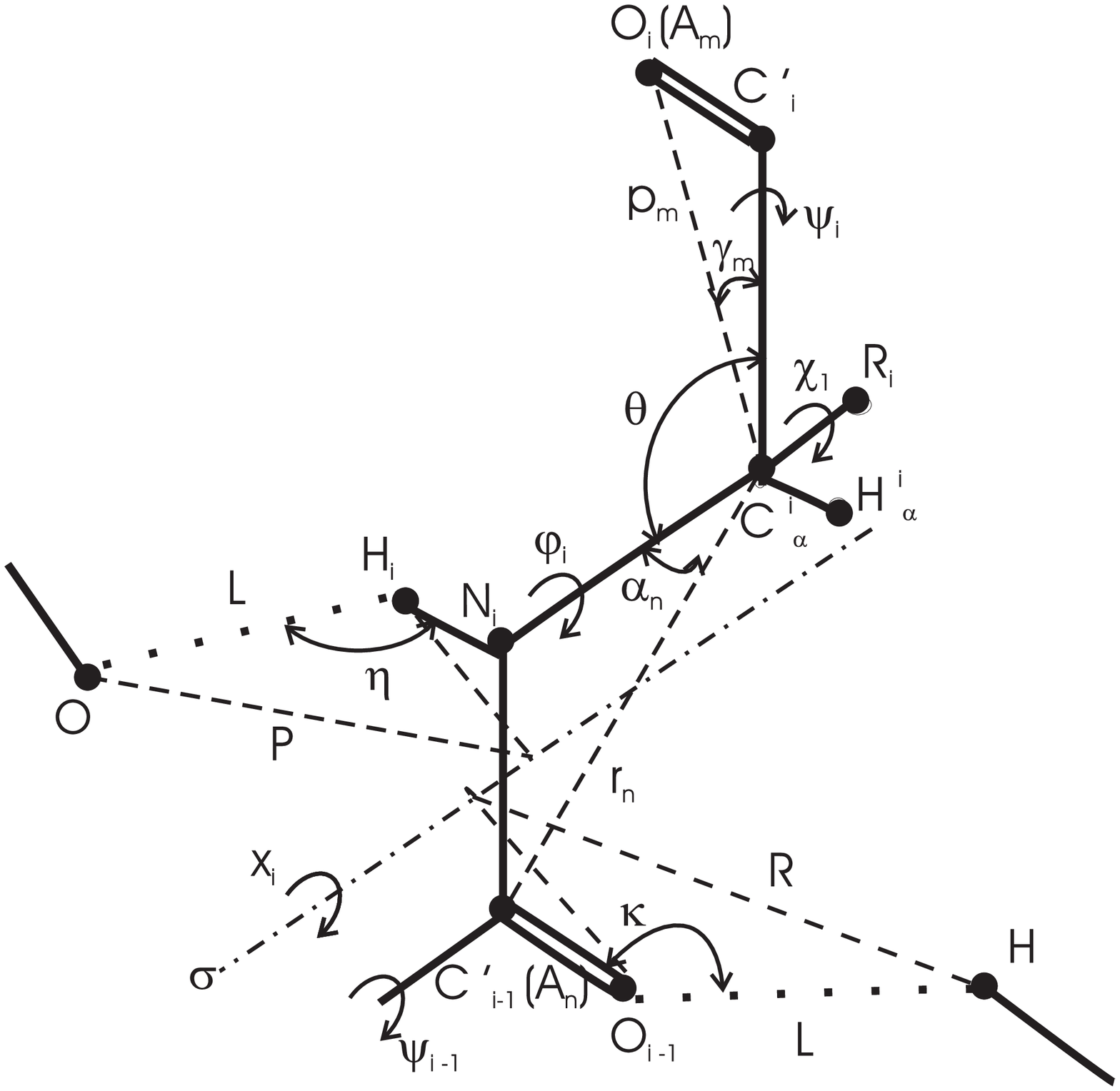}
\end{center}
\caption{A schematic representation of the peptide chain with all
designations necessary for the text. The covalent bonds are shown
as bold lines. Dotted lines denote hydrogen bonds. Dashed lines
denote auxiliary axes.} \label{Fig.1}
\end{figure}

\clearpage
\begin{figure}
\begin{center}
\includegraphics* [width=\textwidth] {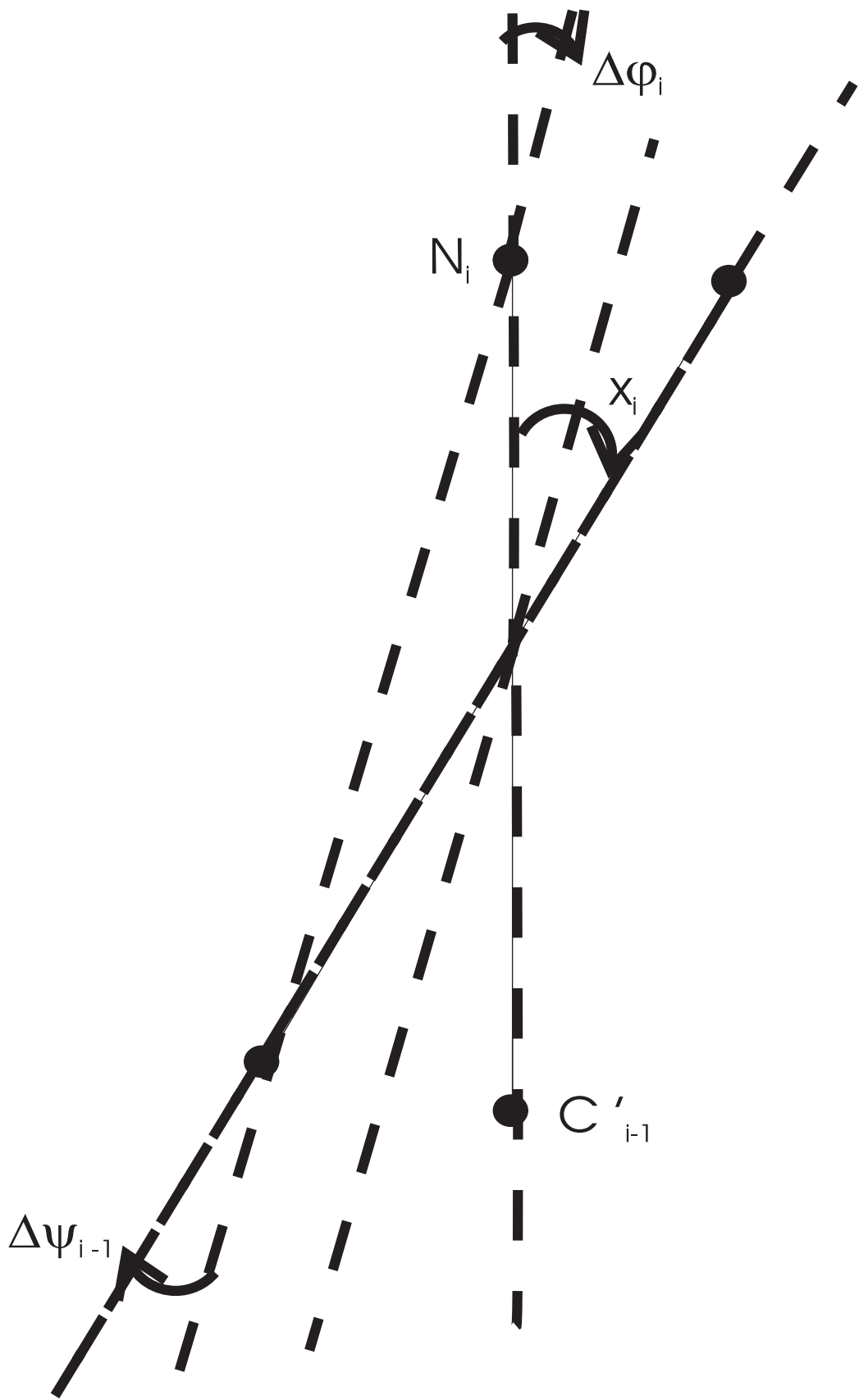}
\end{center}
\caption{A look on the chain from the axis of rotation $\sigma$
explaining the definition of the angle $x_i$ (defined in
(\ref{eq1})).} \label{Fig.2}
\end{figure}

\clearpage
\begin{figure}
\begin{center}
\includegraphics* [width=\textwidth] {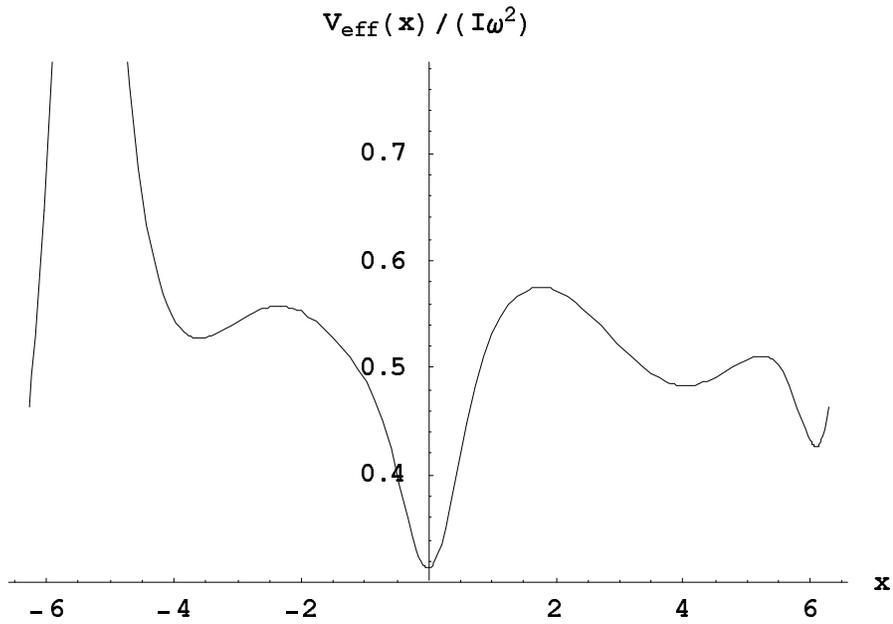}
\end{center}
\caption{The normalized effective on-site potential $V_{eff}(x)$
(see (\ref{eq3})) for the $\alpha $-helix. Here I is the moment of
inertia of the peptide group relative the axis $\sigma$ and
$\omega$ is the frequency. The range for the angular displacement
x (defined in (\ref{eq1})) is 4$\pi$ because it is twice of the
torsional angles $\varphi$ and $\psi$ (see (\ref{eq1})). The part
of the potential over the upper cut off has a maximum with the
value $\approx$ 1.3.} \label{Fig.3}
\end{figure}

\clearpage
\begin{figure}
\begin{center}
\includegraphics* [width=\textwidth] {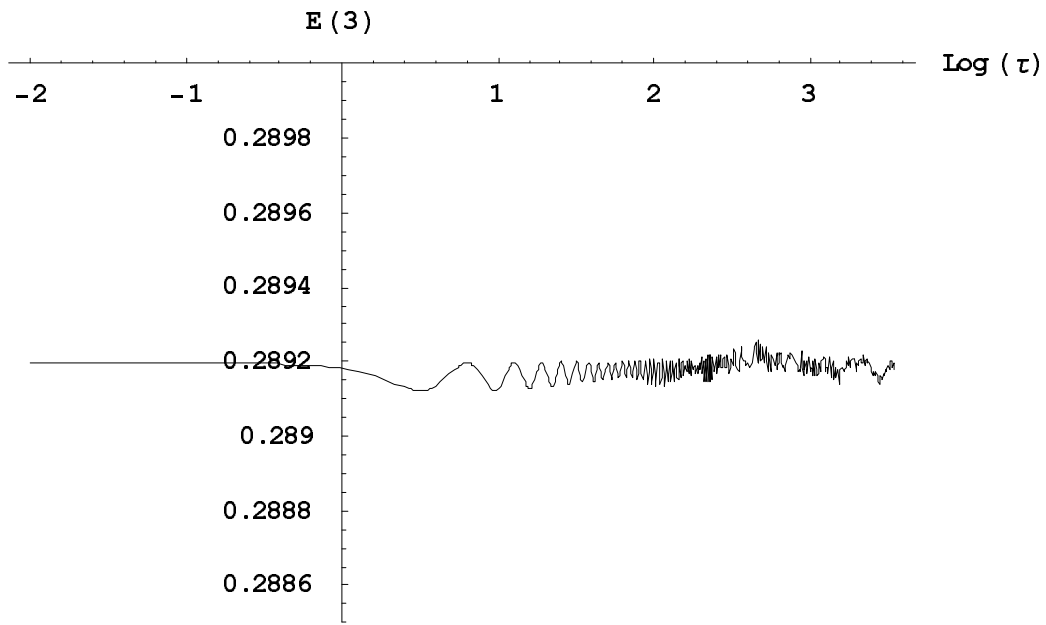}
\end{center}
\caption{The dependence of the energy in the 3-site discrete
breather determined by (\ref{eq21}) and (\ref{eq22}) on
dimensionless time $\tau$ (\ref{eq19}) in the log-time scale for
the $\alpha $-helix.} \label{Fig.4}
\end{figure}

\clearpage
\begin{figure}
\begin{center}
\includegraphics* [width=\textwidth] {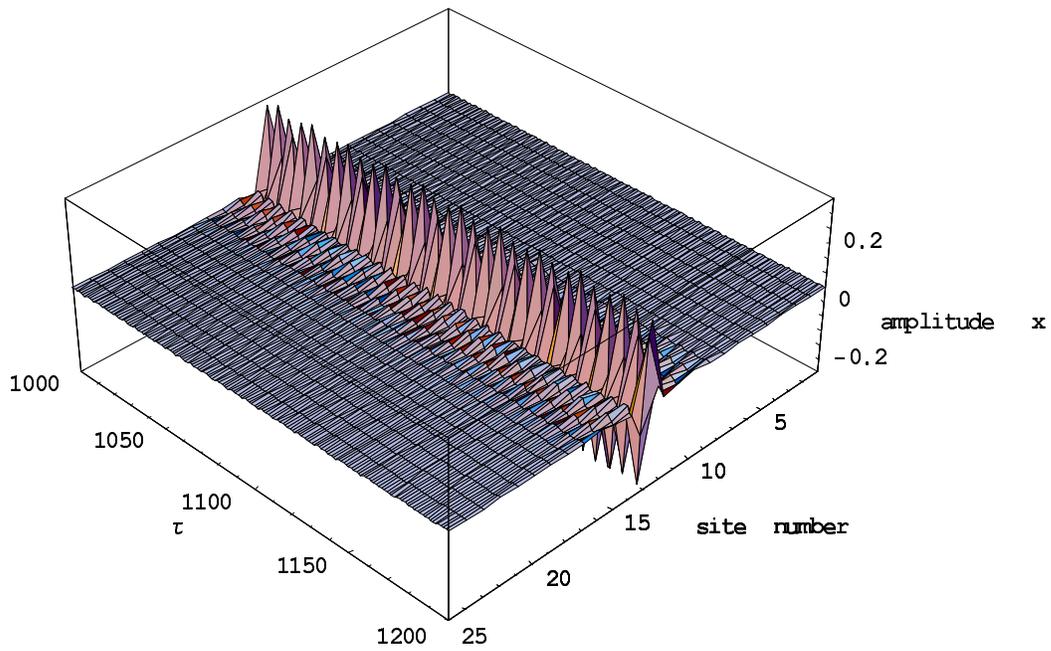}
\end{center}
\caption{The explicit picture of the long time (in dimensionless
units (\ref{eq19})) behavior of the discrete breather in the
$\alpha $-helix of 25 peptide groups named here as sites. The
displacement x (defined in (\ref{eq1})) is the angular deviation
of the peptide group from its equilibrium position.} \label{Fig.5}
\end{figure}
\end{document}